\documentclass[a4paper]{jpconf}%
\usepackage{amssymb}
\usepackage{graphicx}
\usepackage{amsmath}
\usepackage{amsfonts}%
\begin{document}

\title{Non-exponential decay in Quantum Mechanics and Quantum Field Theory}
\author{Francesco Giacosa}

\begin{abstract}
We describe some salient features as well as some recent developments
concerning short-time deviations from the exponential decay law in the context
of Quantum\ Mechanics by using the Lee Hamiltonian approach and Quantum Field
Theory by using relativistic Lagrangians. In particular, the case in which two
decay channels are present is analyzed: the ratio of decay probability
densities, which is a constant equal to the ratio of decay widths in the
exponential limit, shows in general sizable fluctuations which persist also at
long times.

\end{abstract}

\address{Institut f\"{u}r Theoretische Physik, Johann Wolfgang
Goethe - Universit\"{a}t, Max von Laue-Str. 1, D-60438 Frankfurt, Germany}

\ead{giacosa@th.physik.uni-frankfurt.de}

\section{Introduction}

Decays are ubiquitous in Physics: examples are tunnelling processes in quantum
mechanics, very slow decays of nuclei (such as double-$\beta$ decays, lifetime
of about $10^{21}$ y), as well as very fast decays of hadrons and the Higgs
particle (lifetime of about $10^{-22}$ sec). Although the involved decay times
are astonishingly different, the basic phenomenon is the same: a irreversible
transition (infinite Poincare' time) of an unstable initial state coupled to a
continuum of final states.

Both in the context of Quantum Mechanics (QM) and Quantum\ Field Theory (QFT),
an unstable state $\left\vert S\right\rangle $ is not an eingenstate of the
Hamiltonian $H$ of the system, but is described by the so-called energy
distribution $d_{S}(E)$: $d_{S}(E)dE$ is the probability that the unstable
state $\left\vert S\right\rangle $ has an energy between $E$ and $E+dE.$ The
survival probability amplitude $a(t)$ that the state $\left\vert
S\right\rangle $, prepared at the initial time $t=0$, has not yet decayed at
the instant $t>0$, is given by%
\begin{equation}
a(t)=\left\langle S\right\vert e^{-iHt}\left\vert S\right\rangle
=\int_{-\infty}^{+\infty}d_{S}(E)e^{-iEt}dE\text{ }.\label{at}%
\end{equation}
The survival probability is given, as usual, by the amplitude squared:
$p(t)=\left\vert a(t)\right\vert ^{2}.$ The exponential decay $p(t)=e^{-\Gamma
t}$ is obtained when $d_{S}(E)$ takes the Breit-Wigner (BW) form:
$d_{S}^{\text{BW}}(E)=\frac{\Gamma}{2\pi}\left[  (E-M)^{2}+\Gamma
^{2}/4\right]  ^{-1}$. This is however an unphysical limiting case, because a
physical distribution $d_{S}(E)$ should fulfill the following two
requirements: (i) Existence of an energy threshold, $d_{S}(E)=0$ for $E<E_{0}%
$, which suffices to prove that long-time deviations exist, usually implying a
power-law of the type $p(t)\propto t^{-\alpha}$ \cite{khalfin,ghirardi}. (ii)
Finiteness of the mean energy, $\left\langle E\right\rangle =\int_{-\infty
}^{+\infty}Ed_{S}(E)dE<\infty$, which implies that $p^{\prime}(0)=0$ and
therefore that deviations at short times occur
\cite{ghirardi,mercouris,duecan}.

Moreover, a non-exponential behavior at short times implies the existence of
the quantum Zeno effect, which is the inhibition of the decay (or, more in
general, of a quantum transition), due to reiterated frequent measurements
\cite{misra,facchiprl,shimizu}. From an experimental standpoint, the Quantum
Zeno effect was observed by inhibition of a Rabi oscillation between atomic
energy levels \cite{itano}. An improved set-up, in which the Zeno effect was
verified by using single ions and by means of interaction-free repeated
measurements, was shown in Ref. \cite{balzer}.

For what concerns the decay law of a `genuine' unstable quantum state, the
first experimental demonstration of short-time deviations from the exponential
law has been observed by studying the tunneling of cold atoms in an optical
potential \cite{raizen1}. Later on, the same group \cite{raizen2} could verify
also the Zeno (as well as the Anti-Zeno) effect by applying a series of
intermediate measurements on this system. Finally, the experimental proof of
long-time deviations from the exponential law has been observed by
investigating the decay via fluorescence of organic molecules \cite{rothe}.

An experiment performed in the GSI has also found deviations from the
exponential decay law in Hydrogen-like ions decaying via electron capture
\cite{gsianomaly}. In Ref. \cite{gp} these deviations have been linked to a
modification of the BW distribution due to interactions with the measuring
apparatus. Other explanations, based on neutrino oscillations or energy level
splitting of the initial state, have also been put forward \cite{giunti}.

In this proceeding we review some aspects of the non-exponential decay at
short times both in QM and in QFT. In Sec. 2 we present for the QM case a
simple but general Hamiltonian to elucidate the main points. Then, we describe
two recent developments in QM: the broadening of the emitted spectrum at short
times \cite{giacosapra} and the case of a decay into two decay channels
\cite{duecan}. In Sec. 3 we turn to the QFT case and we describe the recent
results of Refs. \cite{duecan,zenoqft,lupofermionico}, in which it has been
shown that, contrary to previous claims \cite{bernardini}, deviations from the
exponential decay law also take place in a genuine QFT framework. Finally, in
Sec. 4 we present our conclusions.

\section{Non-exponential decay in QM}

We introduce a general approach to study decays in QM, which is based on the
so-called Lee Hamiltonian $H=H_{0}+H_{1}$ \cite{lee}:
\begin{equation}
H_{0}=M\left\vert S\right\rangle \left\langle S\right\vert +\int_{-\infty
}^{+\infty}dk\omega(k)\left\vert k\right\rangle \left\langle k\right\vert
\text{ , }H_{1}=\int_{-\infty}^{+\infty}dk\frac{gf(k)}{\sqrt{2\pi}}\left(
\left\vert k\right\rangle \left\langle S\right\vert +\left\vert S\right\rangle
\left\langle k\right\vert \right)  \text{ }.\label{lh}%
\end{equation}
The quantity $g$ is a coupling constant with the dimension of energy$^{1/2}$.
The dimensionless function $f(k)$ describes the interaction of the unstable
state $\left\vert S\right\rangle $ with the `decay-product' state $\left\vert
k\right\rangle $; the energy $\omega(k)$ is the energy of $\left\vert
k\right\rangle $ in the interaction free case. Similar approaches to Eq.
(\ref{lh}) are used in other areas of physics, e.g. in quantum optics with the
Jaynes-Cummings Hamiltonian. The spectral function $d_{S}(E)$ arises as the
imaginary part of the propagator, $d_{S}(E)=\frac{1}{\pi}\operatorname{Im}%
\left[  \left(  E-H+i\varepsilon\right)  ^{-1}\right]  $, and is correctly
normalized, $\int_{-\infty}^{+\infty}d_{S}(E)dE=1$.

The BW limit is obtained for $f(k)=1$ and $\omega(k)=k$. The decay width reads
$\Gamma=g^{2}$ and the spectral function $d_{S}^{\text{BW}}(E)=\frac{\Gamma
}{2\pi}\left[  (E-M)^{2}+\Gamma^{2}/4\right]  ^{-1}$ \cite{giacosapra}. Note,
the choice $\omega(k)=k$ means that the energy is not bounded from below and
the choice $f(k)=1$ means that the unstable state $\left\vert S\right\rangle $
couples to each state of the continuum with the same intensity. For generic
functions $f(k)$ and $\omega(k)$ the decay is not exponential (see below), but
is usually very well approximated by an exponential decay where the decay
width is given by the Fermi golden rule: $\Gamma=g^{2}f^{2}(k_{M})/\left\vert
\omega^{\prime}(k_{M})\right\vert $, where $\omega(k_{M})=M$.

Restricting to the BW limit, an interesting feature is found when studying the
energy of the emitted spectrum. For definiteness, we consider the case of
atomic spontaneous emission: an electron decays from an atomic excited state
to the ground state by emitting a photon. This is indeed the original
framework in which the decay was studied \cite{ww} (see also e.g. Refs.
\cite{scully,facchispont} and refs. therein). Denoting $M$ as the energy
difference between the excited and the ground state, the energy distribution
of the photon at the time $t>0$ is given by $\eta(t,\omega)=\frac{\Gamma}%
{2\pi}\left\vert \frac{e^{-i\omega t}-e^{-i(M-i\Gamma/2)t}}{\omega
-M+i\Gamma/2}\right\vert ^{2}$. The energy uncertainty of the photon
$\delta\omega=\delta\omega(t)$ is given by the width at mid height:
$\eta(t,M)/2=\eta(t,M+\delta\omega/2)$. For $t\rightarrow\infty$ one has
$\delta\omega\rightarrow\Gamma$, which is the natural broadening of the
spectral line. For $t\lesssim3/\Gamma$ the quantity $\delta\omega$ increases
as $\delta\omega\simeq5.56/t$. This means that, if measured at short times
after the emission, the photon can have an energy which deviates from the mean
value $M$ by several decay widths $\Gamma$, see Ref. \cite{giacosapra} for details.%

\begin{figure}
[ptb]
\begin{center}
\includegraphics[
height=2.0903in,
width=5.4708in
]%
{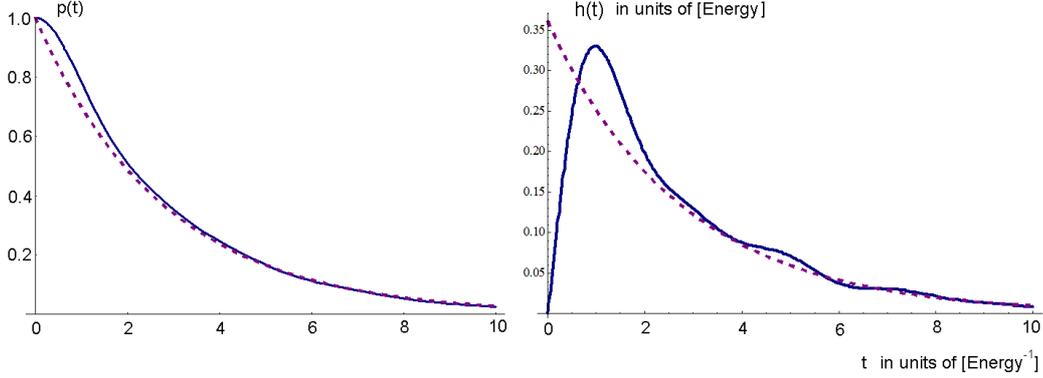}%
\caption{The functions $p(t)$ (left panel, solid line) and $h(t)=-p^{\prime
}(t)$ (right panel, solid line) are shown for $f(k)=\theta(k-E_{0}%
)\theta(\Lambda-k)$ with (arbitrary energy units): $E_{0}=0$, $\Lambda=5,$
$M=2,$ $g^{2}=0.36$. The (exponential) dashed lines correspond to the BW
limit.}%
\end{center}
\end{figure}

We now turn to the a simple example of non-exponential decay. To this end, we
choose $f(k)=\theta(k-E_{0})\theta(\Lambda-k),$ which means that the state
$\left\vert S\right\rangle $ couples \emph{only }to an energy window of final
states. A numerical example is shown in Fig. 1, in which the survival
probability $p(t)$ and its derivative, $h(t)=-p^{\prime}(t)$, are plotted
(solid lines) and compared with the exponential counterparts (dashed lines,
obtained by using the Fermi rule). Deviations from the exponential decay are
evident. Note, the function $h(t)$ has a clear physical meaning:
$h(t)dt=p(t)-p(t+dt)$ is the probability that the decay occurs in the
time-interval $(t,t+dt)$. For this reason, $h(t)$ is also denoted as the
density of decay probability.%

\begin{figure}
[ptb]
\begin{center}
\includegraphics[
height=2.3255in,
width=3.2811in
]%
{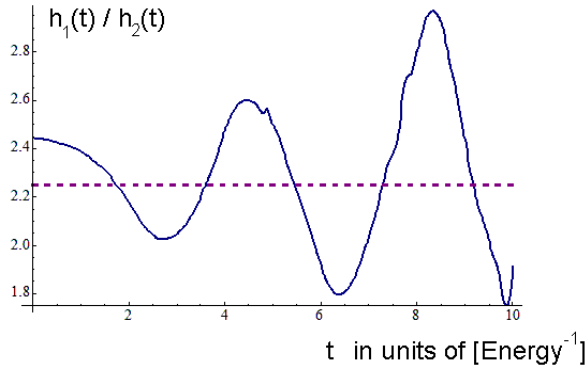}%
\caption{The solid curve is the ratio $h_{1}(t)/h_{2}(t)$ plotted as function
of $t$ for the numerical choice (arbitrary energy units) $E_{0,1}=0,$
$E_{0,2}=0.5,$ $\Lambda=5,$ $g_{1}^{2}=0.36,$ $g_{2}^{2}=0.16.$ The constant
dashed line corresponds to the BW limit $\Gamma_{1}/\Gamma_{2}$. }%
\end{center}
\end{figure}

We now turn to the two-channel case, in which the unstable state $\left\vert
S\right\rangle $ couples to two sets of final states $\left\vert
k,1\right\rangle $ and $\left\vert k,2\right\rangle .$ The Hamiltonian is a
straightforward generalization of Eq. (\ref{lh}):%

\[
H_{0}=M\left\vert S\right\rangle \left\langle S\right\vert +\sum_{i=1,2}%
\int_{-\infty}^{+\infty}dk\omega_{i}(k)\left\vert k,i\right\rangle
\left\langle k,i\right\vert \text{ , }H_{1}=\sum_{i=1,2}\int_{-\infty
}^{+\infty}dk\frac{g_{i}f_{i}(k)}{\sqrt{2\pi}}\left(  \left\vert
S\right\rangle \left\langle k,i\right\vert +\text{h.c.}\right)  \text{ .}
\]
In the BW limit $f_{i}(k)=1$ and $\omega_{i}(k)=k+\alpha_{i},$ where
$\alpha_{1}$ and $\alpha_{2}$ are constants. The partial decay widths are
$\Gamma_{1}=g_{1}^{2}$ and $\Gamma_{2}=g_{2}^{2}$ and the survival probability
reads $p(t)=e^{-\Gamma t}$ with $\Gamma=\Gamma_{1}+\Gamma_{2}$.

In the presence of two decay channels, it is useful to introduce the quantity
$h_{i}(t)$: $h_{i}(t)dt$ is the probability that the state $\left\vert
S\right\rangle $ decays in the $i$-th channel between $t$ and $t+dt.$ In the
BW limit $h_{i}(t)=\Gamma_{i}e^{-\Gamma t}$ and the ratio $h_{1}(t)/h_{2}(t)$
is a constant equal to $\Gamma_{1}/\Gamma_{2}$. However, when deviations from
the exponential decay are considered, the ratio $h_{1}(t)/h_{2}(t)$ is in
general not a constant, but shows sizable departures from $\Gamma_{1}%
/\Gamma_{2}$ \cite{duecan}. This is shown for a particular numerical example
in Fig. 2, in which the choices $f_{i}(k)=\theta(k-E_{0,i})\theta(\Lambda-k)$
and $\alpha_{i}=0$ have been used.

\section{Non-exponential decay in QFT}

QFT is the theoretical framework in which particles are created and
annihilated. It is therefore the most fundamental environment to study the
nature of decays. In Refs. \cite{bernardini} it was claimed that in QFT the
deviations from the exponential decay law do not take place. These
conclusions, however, rely on a perturbative treatment of the decay law. This
issue was (re)analyzed in Refs. \cite{duecan,zenoqft} with the following results:

\begin{enumerate}
\item In Ref. \cite{duecan,zenoqft,achasov,lupo} the superrenormalizable QFT
interaction Lagrangian $\mathcal{L}=gS\varphi^{2}$, leading to the decay
$S\rightarrow\varphi\varphi$, was studied in detail. The spectral function
$d_{S}(E)$ is calculated as the imaginary part of the propagator of $S.$ To
this end, it is necessary to perform a resummation of (at least) the one-loop
self-energy diagram of the field $S$. (Namely, a perturbative expression of
$d_{S}(E)$ in a series of $g$ is not a meaningful quantity \cite{duecan}.) The
survival probability takes the very same formal expression of Eq.
(\ref{at})and short-time deviations from the exponential law do occur
\cite{duecan,zenoqft}. Quite remarkably, the time interval in which such
deviations take place is independent on the cutoff \cite{zenoqft}. This is due
to the fact that the energy distribution $d_{S}(E)$ behaves as $E^{-3}$ for
large energies, thus being insensitive on the high energy scale of the theory.

The two-channel case can be easily studied by using the Lagrangian
$\mathcal{L}=g_{1}S\varphi_{1}^{2}+g_{2}S\varphi_{2}^{2},$ which implies that
the two decay processes $S\rightarrow\varphi_{1}\varphi_{1}$ and
$S\rightarrow\varphi_{2}\varphi_{2}$ take place. The ratio $h_{1}(t)/h_{2}(t)$
shows also in this QFT framework qualitatively similar deviations from the
constant BW limit as those in\ Fig. 2, see Ref. \cite{duecan} for plots and details.

\item In Ref. \cite{lupofermionico} the renormalizable interaction Lagrangian
$\mathcal{L}=gS\bar{\psi}\psi,$ where $\psi$ is a fermionic field, has been
analyzed. (Note, the Higgs coupling to fermions has the same form.) The energy
distribution $d_{S}(E)$ scales only as $E^{-1},$ which in turn implies that
the presence of a high energy cutoff is necessary for its correct
normalization. Thus, the existence of a high energy cutoff is not only a
mathematical step, but is a necessary physical requirement. In this framework,
the duration of the deviations from the exponential decay at short times lasts
only $\Lambda^{-1},$ and is thus very small if the high-energy cutoff
$\Lambda$ is large.
\end{enumerate}

\section{Summary and outlook}

In this proceeding we have discussed the deviations from the exponential decay
law in QM and QFT. The decay law in\ QM has been studied by using the Lee
Hamiltonian approach \cite{duecan,giacosapra} and in QFT by using relativistic
QFT Lagrangians \cite{duecan,zenoqft,lupofermionico}.

In QM we have discussed a peculiar property of the energy spectrum of the
photon in a spontaneous emission process: even in the exponential limit, a
broadening of the emitted spectrum is realized if the photon is measured at
early times \cite{giacosapra}. As a next step we have studied short-time
deviations from the exponential law in the simple case in which the unstable
state couples to the continuum in a window of energy (see Fig. 1 and Refs.
\cite{duecan,giacosapra}).

A particularly interesting feature in both QM and QFT is that of a decay in
which two (or more) decay channels are present. The ratio of decay probability
densities is a constant in the BW limit and equals the ratio of decay widths,
but show sizable fluctuations on top of this constant limit when deviations
from the BW limit are included (see Fig. 2 and Ref. \cite{duecan}). A
systematic study of this ratio for a variety of cases, also including the
analysis of realistic physical situations, is an important outlook for the future.

\bigskip

\textbf{Acknowledgments: } the author thanks G. Pagliara and T.
Wolkanowski-Gans for useful discussions and the Foundation of the
Polytechnical Society of Frankfurt for support through an Educator fellowship.

\end{document}